# Empirical Evidence on Genre-Time Correlation in Box-Office Success Using Exploratory Data Analysis and Machine Learning

**Iqra Tariq**
Iqra.tariq@gift.edu.pk

**ABSTRACT**
The movie industry is one of the fastest-growing global sectors, characterised by high production costs and significant financial risk. Given the capital-intensive nature of filmmaking, accurately predicting box office success is of critical importance for stakeholders ranging from producers to investors. This study investigates the correlation between movie genre and release timing as predictive factors for commercial success. A combined approach involving Exploratory Data Analysis (EDA) and supervised machine learning techniques is proposed to assess this relationship. The dataset, comprising the top 200 box office hits and the top 100 flops, was curated from reliable sources, including IMDb, Box Office Mojo, The Numbers, and Wikipedia. EDA revealed that specific genres show statistically significant patterns of success or failure in particular months. For instance, animated and superhero movies achieved their peak success rates in June and July (28% and 29%, respectively), while thrillers and romance genres showed higher hit rates in November. Conversely, the flop dataset showed genres like action and comedy more frequently underperforming in March, April, and August. To validate these findings, multiple regression-based machine learning models were applied using both cross-validation (5-, 10-, and 15-fold) and percentage-split methods (60–40, 70–30, and 80–20). Algorithms such as LWT, Multilayer Perceptron, Random Tree, and Decision Stamp demonstrated high predictive accuracy, reinforcing the hypothesis of genre-time dependency. The results consistently indicated a strong correlation between release month and genre performance, providing valuable insight for strategic planning in content production and release scheduling. This study highlights the growing need to apply data analytics in the media industry, like other data-driven domains, for risk mitigation and optimized decision-making. Future work may expand on cross-industry comparisons and examine variations in genre-time correlation between OTT platforms and theatrical releases.

*Note: This paper is an empirical case study of a large movie analytics research work to examining the **correlation between genre x month** that effect box office performance. Initial results are calculated on 300 movies dataset, (global 200 hit & 100 Flop) using EDA and pretrained regression models.

# Empirical Evidence on Genre-Time Correlation in Box-Office Success Using Exploratory Data Analysis and Machine Learning

**Iqra Tariq**
Iqra.tariq@gift.edu.pk

**ABSTRACT**
The movie industry is one of the fastest-growing global sectors, characterised by high production costs and significant financial risk. Given the capital-intensive nature of filmmaking, accurately predicting box office success is of critical importance for stakeholders ranging from producers to investors. This study investigates the correlation between movie genre and release timing as predictive factors for commercial success. A combined approach involving Exploratory Data Analysis (EDA) and supervised machine learning techniques is proposed to assess this relationship. The dataset, comprising the top 200 box office hits and the top 100 flops, was curated from reliable sources, including IMDb, Box Office Mojo, The Numbers, and Wikipedia. EDA revealed that specific genres show statistically significant patterns of success or failure in particular months. For instance, animated and superhero movies achieved their peak success rates in June and July (28% and 29%, respectively), while thrillers and romance genres showed higher hit rates in November. Conversely, the flop dataset showed genres like action and comedy more frequently underperforming in March, April, and August. To validate these findings, multiple regression-based machine learning models were applied using both cross-validation (5-, 10-, and 15-fold) and percentage-split methods (60–40, 70–30, and 80–20). Algorithms such as LWT, Multilayer Perceptron, Random Tree, and Decision Stamp demonstrated high predictive accuracy, reinforcing the hypothesis of genre-time dependency. The results consistently indicated a strong correlation between release month and genre performance, providing valuable insight for strategic planning in content production and release scheduling. This study highlights the growing need to apply data analytics in the media industry, like other data-driven domains, for risk mitigation and optimized decision-making. Future work may expand on cross-industry comparisons and examine variations in genre-time correlation between OTT platforms and theatrical releases.

## INTRODUCTION

The world has undergone transformations that were previously unimaginable due to the emergence of big data and analytics(Bryda & Costa, 2023; Ahamed & Sridevi., 2025). Data analytics has created numerous vital new pathways that can be utilized to examine previous data, make informed marketing decisions, and accurately anticipate the fortunes of any product in its relevant market and industry(Bryda & Costa, 2023; Brewis et al., 2023)]. Like many other top-grossing industries, the motion picture industry is rapidly developing and thriving (Cuntz et al., 2024). Millions of dollars are invested annually in the development and production of movies, and a considerable group of people are involved in this business to make the movie a complete package (Cuntz et al., 2024; Memon & Hussain, 2025). Some people directly represent the face of cinema, i.e., male and female actors, while others become part of movie development, pre-production, and post-production teams (Mykhailiukova & Antonivska, 2024). Focusing on the importance of this major money-making industry in the GDP growth of multiple countries, numerous machine learning methods have been introduced and used to predict, estimate, and calculate the success of a movie(Weber et al., 2024; Kristianto et al., 2024).

The diversity of the audience and their secondary differences (Geographical region, language, culture, etc.) make the acceptance of a movie more complex (Tiwary, 2024). There are multiple factors involved in the overall success, including box-office collection and content popularity, which encompass story, Actors, Computer-Generated Imagery (CGI), Production Capital, Entertaining factors (such as music and humor), Trends, Word-of-Mouth, etc.(Memon & Hussain, 2025; ; Kristianto et al., 2024; Bsoul et al., 2025; Qiao, 2024; Xianxian & Shafii, 2024; Aondover, 2025; Kumar et al.,2024) .

On the other hand, despite focusing on and considering all the important factors for the success of a movie, some movies still fail to shine at the box office, making a big dent on the production capital (Memon & Hussain, 2025; Darapaneni, et al., 2020). Table 1 shows the top 10 movies with the highest production budgets that failed to achieve box office success in the market.

Table 1. List of movies with big budget and flop status

| No. | Movie | Approx. Income | Approx. Expense | Profit |
|---|---|---|---|---|
| 1 | Turning Red | $7,625,620 | $175,000,000 | -167374380.00 |
| 2 | Jungle Cruise | $113,816,890 | $264,216,000 | -150399110.00 |
| 3 | Mars Needs Moms | $26,775,915 | $170,166,000 | -143390085.00 |
| 4 | Mulan | $59,259,506 | $200,000,000 | -140740494.00 |
| 5 | Moonfall | $32,168,901 | $170,784,000 | -138615099.00 |
| 6 | Wonder Woman 1984 | $122,262,002 | $258,940,000 | -136677998.00 |
| 7 | The Matrix Resurrections | $85,905,124 | $218,710,000 | -132804876.00 |
| 8 | Onward | $86,942,784 | $217,240,000 | -130297216.00 |
| 9 | The Suicide Squad | $99,178,304 | $218,942,000 | -119763696.00 |
| 10 | John Carter | $183,200,614 | $295,824,000 | -112623386.00 |

## *Motivation Of The Study*

As researchers, one of our primary goals is to investigate the business side of the movie industry and examine the economics that underlie what makes a movie commercially successful. We are primarily interested in determining; i) whether any patterns exist among films that cause them to become successful at the box office and ii) whether the success of a film's genre at a particular time of the year correlates with that film's box office success. If we had access to a helpful analysis, we would be able to forecast a movie's box office success before it premieres, without relying on reviewers' opinions. More precisely, the primary focus of this research is to identify any patterns related to movie success at a particular time of the year before the marketing and release phase, specifically concerning genre or movie category. Previously movie release is done on the bases of seasonality and leaving the gap of monthly impact on the box office.

## *Significance Of The Intended Research*

The popularity of a movie is not solely dependent on features that are associated with the it. The size of the audience is one of the most critical factors that determines whether a film is successful. If there is no audience to watch a movie, then the entire industry is pointless. For the provided argument, it is very important to schedule the release of the film to the respective audience origin and genre likeness. We carry out an exploratory study of the data and discover some fascinating phenomena. Therefore, to support our argument, we have also applied machine learning algorithms on movie data to identify the relation among these important attributes (Month and Genre) for the prediction of movie success before its release. This not only enables us to learn about the factors that contribute to a movie's success at the box office, but it also helps us verify our claim.

## Research Workflow

To find the link between genre and time, the following workflow has been adopted. The phases elaboration (Table 2) shows all the relevant activities associated with each phase along with the resources and tools used during the process.

Table 2 Research Framework

| Index | Phase | Activities | Sources/Tool |
|---|---|---|---|
| 1 | Data Collection | Data gathering according to the shortlisted attributes | IMDB. NUMBERS, IMDB Mojo, Wikipedia |
| 2 | Data Preprocessin g | Data Cleaning, Data Conversion from textual to Numerical (Supervised). Data Distribution. | MS. Excel |
| 3 | EDA | Genre-based movies' success ratio with respect to all movies data and time trends and patterns. | MS. Excel |
| 4 | Purposed Attributes | | |
| 5 | Machine Learning | Classification process for the model training and testing with Eleven different classifiers. | Weka Software |
| 6 | Comparativ e Data Analysis | Comparative analysis based on Five performance measures (CC, MAE, RMSE, RAE, RRSE). | MS. Excel |
| 7 | Discussion | | |

### *Contribution*

The contribution of our work is summarized as follows

1. We have performed EDA on the Movies dataset (Flop and Hit) to find the correlation between Genre-Time concerning the top fourteen genres over twelve months of the year.
2. We have identified the month-wise best and worst periods of the year to release a particular genre.
3. We have performed comparative analysis to verify the EDA results on the basis of cross-fold validation over three cross folds (5,10, and 15) as well as the Percentage split approach (60,70, and 80) on the dataset for regression analysis on the dataset of hit and flop movies.
4. We have proposed the importance of Time in terms of months as an attribute to calculate and predict the success of a particular Genre release.

## LITERATURE REVIEW

Time series analysis has emerged as a valuable tool for identifying patterns and trends in genre popularity over time. [16] emphasize that by analysing historical data, it is possible to predict future genre popularity and its impact on box office success. This perspective acknowledges that a movie's success is not determined solely by its inherent qualities but also by the prevailing cultural trends and audience preferences at the time of release (Shahid et al., 2023).

Genre popularity plays a significant role in influencing box office performance. So, Arora and Rani (2024) also note that time series analysis is successful in predicting genre trends and that doing so as an additional feature for movie success prediction models could lead to more accurate movie success prediction models (Arora & Rani, 2024). Zhang et al. (2025) supports the claim by finding evidence that time series-based `genre popularity' features are useful for predicting the success of movies (Zhang et al., 2025). It also emphasizes the importance of release time in terms of a movie's profitability (Memon & Hussain, 2025; Singh & Rokde; 2024).

Historically, genre popularity has been treated as a static quantity by traditional methodologies. According to Shahid and Islam (2021), it is novel and useful to use time series techniques to estimate the popularity of genres (Shahid & Islam, 2021). Prediction models that include time factors learn from temporal trends and adjust their predictions to better represent future behaviours. Hussain (2022) also suggests that time series- based features can capture the popularity of a story plot and such practical features can improve accuracy in predicting movie success (Shahid et al., 2023).

The difficulty of predicting the success of a movie prior to release has caused the film industry to increasingly employ machine learning for the reasons like natural language processing for sentiment analysis. Xie (2024) While it is recognized that these models make use of pre-release information to predict the box office receipts that will fuel marketing actions Xie, 2024). Another advantage of early prediction is provided by Shahid and Islam (2021), who argue that it enables making decision on investments (Shahid et al., 2023). By considering which genres are popular, filmmakers and investors can make more informed decisions when deciding which project to pursue, this could facilitate more efficient allocation of resources and a greater chance of commercial success (Shahid et al., 2023).

Genre-based classification systems employing machine learning and clustering methods are also adapted to predict box-office performance. Singh et al. (2024) argue that, coupling such systems to tools for time series analysis, leads to a holistic view of the factors affecting the success of a movie (Singh et al., 2024). Such a rich dataset that includes genre, cast, crew, budget, release timing can be utilized in ML models (Singh et al., 2024).

By combining time series analysis with machine learning we have arrived at a very powerful feature set for predicting movie success. Whilst time series analysis focuses on the temporal dynamics of genre popularity, machine learning models enable us to further consider other covariates of interest, such as cast, budget and marketing (Shahid et al., 2023; Singh et al., 2024). The approach which is based on the combination of these methods is expected to provide more accurate and robust predictions.

## MATERIAL AND METHODS

### *Collection Of Data*

To pursue the targeting milestone regarding the movie success prediction data has been collected from three main sources, IMDB, IMDB Mojo, and Wikipedia. These three have all provided the necessary information, which has been compiled for the final prediction.

The following are the main sources, and the respected data collected from each source is mentioned alongside.

1. We have used IMDB Mojo for collecting facts of the top 200 all-time box office hit and top 100 box office disaster movies with dimensionality of 4 variables (Title, Genre, Box Office, Release Month). This data also has been used for descriptive analysis based on genre-time correlation to box office hit ratio, as well as for the final prediction.
2. We have used Wikipedia official page of top 200 all-time box office hit movies and top 100 box office disaster movies to get the release data in Day-Month format. This data has been used for final prediction with respect to genre-time correlation for movies success prediction.

### *Data Preprocessing (Cleaning And Formatting)*

The data collected was in three different formats. One of the main barriers was to convert the data into the desired format for the final analysis and prediction. For this purpose, following pre-processing operations have been performed on the collected data considering dataset format for selected classifiers.

1. All genre classes were assigned a specific numeric value.
2. Movies titles were represented by numeric identifiers.
3. Month of the movie release was used as numerical value.
4. Status was assigned to all movies (hit=1, flop=0).

### *Exploratory Data Analysis (EDA)*

For EDA top 200 box office hit and 100 flop movies data has been collected with four attributes (Title, Genre, Box Office, Release Month). The data was split based on the genre-month to find whether there exist any patterns for specific genre success rate over different months of the year. As well as in EDA it is evaluated which movies have the greatest number of box office hits over the history of IMDB dataset.

### *Purposed Data Dimensions*

There are many aspects, both large and small, that contribute to the overall success of the movie. Based on EDA following proposed features are being selected for the prediction of movie success in the research. Table 3 provides an explanation of the factors that monitored and considered during the machine learning process.

*Table 3 Final Attributes of Research*

| No. | Feature | Description |
|---|---|---|
| 1 | Index/title | Name of the Movie |
| 2 | Genre | Category of the Movie |
| 3 | Release Data | Release Month |
| 4 | Box Office Record | Total Earning |
| 5 | Status | Success/Failure |

### *Machine Learning for Success Prediction*

Data is converted into numerical values for the final processing in the classification and prediction so that focusing on the final data type and nature of the proposed solution, where Genre-Time correlates we have selected 13 regression classifiers for machine learning model training, Testing and validation.

### *Comparative Analysis*

Tabular and line graph representation of machine learning results have been represented in comparative manor to show the best of the top five classifiers in all three respective data distribution methods for model training., testing and validation.

## RESULTS AND ANALYSIS

### *Genre-Time Correlation by Exploratory Data Analysis*

The term "Exploratory Data Analysis" refers to the crucial process of performing pre-liminary investigations on data to discover patterns, identify outliers, test hypotheses, and verify assumptions with the aid of summary statistics and graphical representations. Considering the purpose, to find the patterns from IMDB full dataset an EDA has been conducted to identify the correlation between Genre and Time of movie release to get the insight about the possible association. The first part of EDA contains an assessment of the number of box office hits from the top 200 movies with respect to genre, Figure 1 shows the relevant statistics over 130 years. Second part of EDA contains the assessment of top 200 box office hits with respect to Genre-Time relationship. Figure 2 shows the number of genre hits in particular month of the year, which indicates a clear insight into the percentage of specific genres in different months.

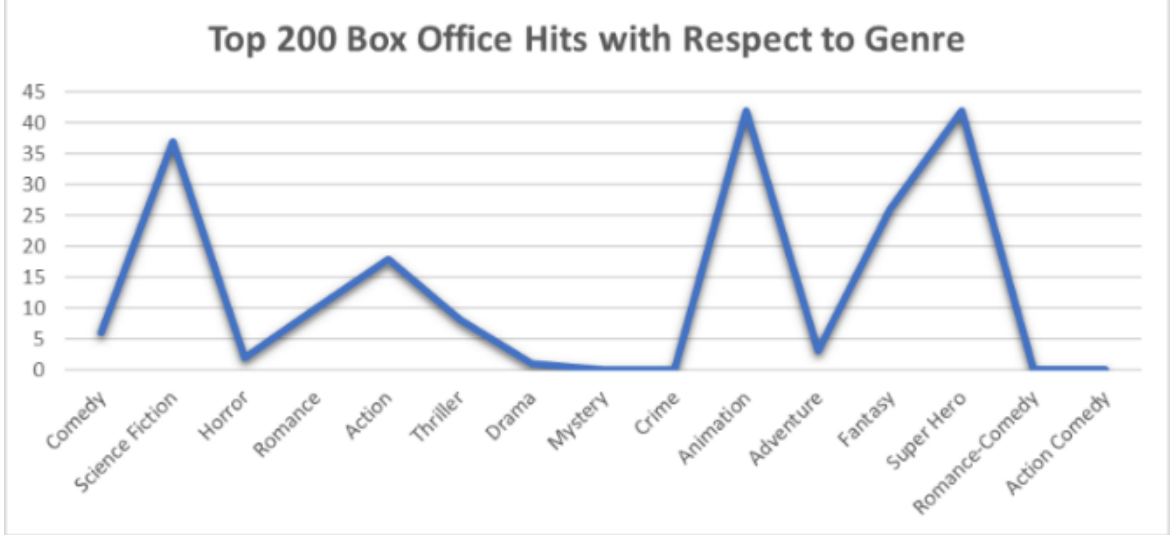

*Figure 1 Genre Success Frequency from the Data*

During the first EDA assessment, it has been observed that most popular genre among mass is Animation and Super Hero with highest number of box office hits (42) over 130 years, followed by and Science Fiction (37) and Fantasy (26) respectively. Moreover, no box office hits took place from crime and mystery genre in top 200 list.

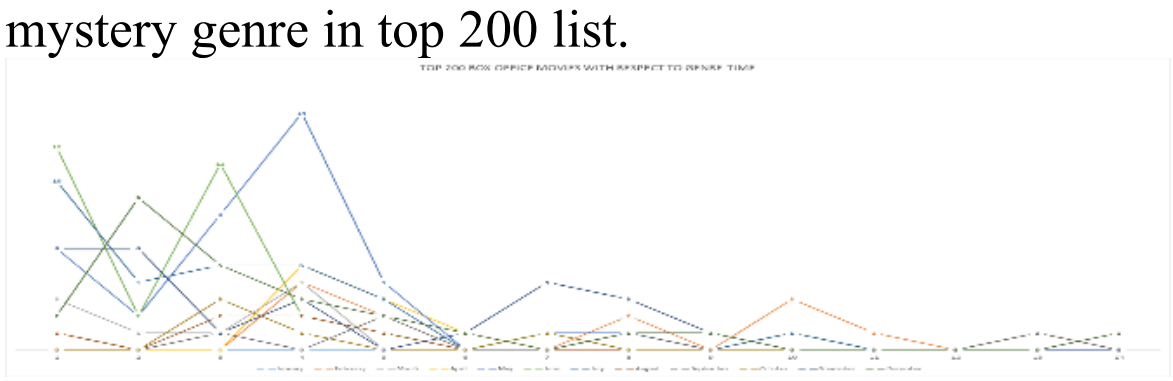

*Figure 2 Genre Success in Particular Month of the Year*

During the second EDA assessment all 200 movies have been graphed and visualized on the bases on the release month. Moreover, individually, movies data has represented in months' respective format to know the occurrence of most promising genre in specific month. Following table shows the assigned categories from the 200 movies list. Genres of top movies were assigned numerical class for the final observation if the patterns, Table 4 shows the assigned classes of respective categories.

*Table 4 Genre Tagging*

| Genre | Class | Genre | Class | Genre | Class |
|---|---|---|---|---|---|
| Animated | 1 | Adventure | 6 | Drama | 11 |
| Fantasy | 2 | Thriller | 7 | Sports | 12 |
| Science Fiction | 3 | Romance | 8 | Supernatural | 13 |
| Superhero | 4 | Biographical | 9 | Monster | 14 |
| Action | 5 | Comedy | 10 | | |

From the data it has been observed that genre has diverse occurrence of success rate over the months of the year. Following graphs indicates the month wise genre hit statistics presented in Figure 3.

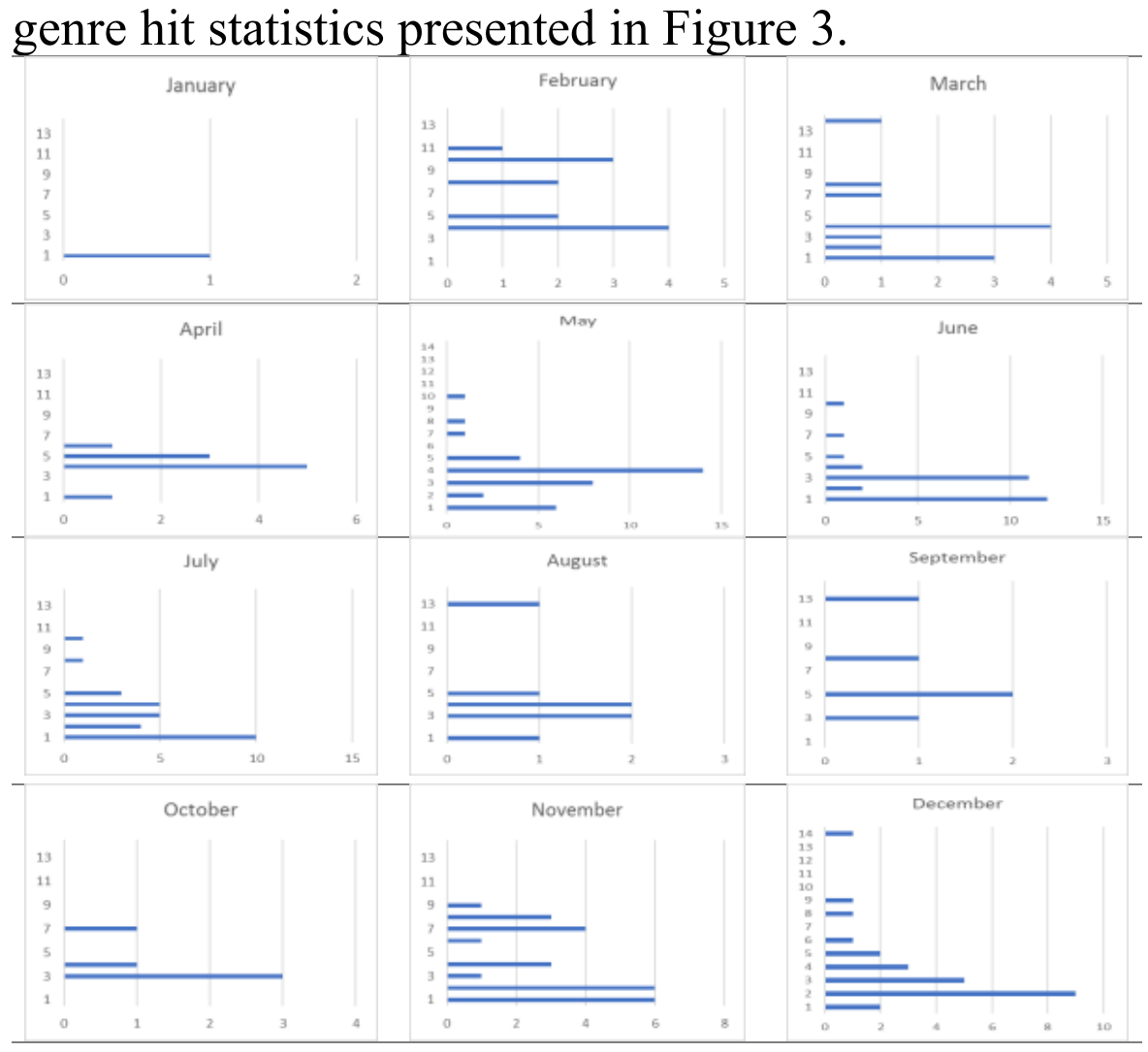

*Figure 3 Month wise Genre Success Rate*

*Table 5 Heatmap of the Genre-Time Relation of the top 200 Box Office hits*

| Class | January | February | March | April | May | June | July | August | September | October | November | December |
|---|---|---|---|---|---|---|---|---|---|---|---|---|
| 1 | 1 | 0 | 3 | 1 | 6 | 12 | 10 | 1 | 0 | 0 | 6 | 2 |
| 2 | 0 | 0 | 1 | 0 | 4 | 2 | 4 | 0 | 0 | 0 | 6 | 9 |
| 3 | 0 | 0 | 1 | 0 | 8 | 11 | 5 | 2 | 1 | 3 | 1 | 5 |
| 4 | 0 | 3 | 3 | 5 | 14 | 2 | 5 | 2 | 0 | 1 | 3 | 3 |
| 5 | 0 | 2 | 0 | 3 | 4 | 1 | 3 | 1 | 2 | 0 | 0 | 2 |
| 6 | 0 | 0 | 0 | 1 | 0 | 0 | 0 | 0 | 0 | 0 | 1 | 1 |
| 7 | 0 | 0 | 1 | 0 | 1 | 0 | 0 | 0 | 0 | 1 | 5 | 1 |
| 8 | 0 | 2 | 1 | 0 | 1 | 0 | 1 | 0 | 1 | 0 | 3 | 1 |
| 9 | 0 | 0 | 0 | 0 | 0 | 0 | 0 | 0 | 0 | 0 | 1 | 1 |
| 10 | 0 | 3 | 0 | 0 | 1 | 1 | 1 | 0 | 0 | 0 | 0 | 0 |
| 11 | 0 | 1 | 0 | 0 | 0 | 0 | 0 | 0 | 0 | 0 | 0 | 0 |
| 12 | 0 | 0 | 0 | 0 | 0 | 0 | 0 | 0 | 0 | 0 | 0 | 0 |
| 13 | 0 | 0 | 0 | 0 | 0 | 0 | 0 | 1 | 1 | 0 | 0 | 0 |
| 14 | 0 | 0 | 1 | 0 | 0 | 0 | 0 | 0 | 0 | 0 | 0 | 1 |

From the above heatmap (Table 5), the following observations can be clearly seen from the Hit movies data.

1. Class 1 has the highest tendency of hit in the month of June and July with 28% rate.
2. Class 2 has a high tendency of hits in the month of December with 28% rate.
3. Class 3 has a high tendency of hits in the month of May and June with 34% rate.
4. Class 4 has a high tendency of hits in the month of May with 29% rate.
5. Class 5 has a high tendency of hits in the month of May with 21% rate.
6. Class 7 has a high tendency of hits in the month of November with 55% rate.
7. Class 8 has high tendency of hits in the month of November with 33% rate.
8. Class 10 has a high tendency of hits in the month of February with 60% rate.
9. Class 6,9,11,12,13 and 14 have least contribution in the box office hit list.

*Table 6 Heatmap of the Genre-Time Relation of top 100 Box Office Flops*

| Class | January | February | March | April | May | June | July | August | September | October | November | December |
|---|---|---|---|---|---|---|---|---|---|---|---|---|
| 1 | 0 | 0 | 3 | 1 | 0 | 1 | 1 | 0 | 0 | 0 | 3 | 1 |
| 2 | 0 | 3 | 2 | 1 | 1 | 3 | 0 | 0 | 1 | 1 | 1 | 5 |
| 3 | 1 | 4 | 1 | 0 | 1 | 2 | 4 | 0 | 3 | 1 | 1 | 1 |
| 4 | 0 | 0 | 0 | 0 | 0 | 3 | 1 | 4 | 0 | 0 | 0 | 0 |
| 5 | 2 | 1 | 0 | 5 | 1 | 1 | 1 | 2 | 1 | 1 | 1 | 0 |
| 6 | 0 | 0 | 0 | 2 | 1 | 0 | 0 | 0 | 0 | 0 | 0 | 0 |
| 7 | 0 | 0 | 0 | 0 | 0 | 1 | 0 | 0 | 0 | 0 | 0 | 0 |
| 8 | 0 | 0 | 0 | 1 | 0 | 0 | 0 | 1 | 1 | 0 | 0 | 0 |
| 9 | 0 | 0 | 0 | 0 | 0 | 0 | 0 | 0 | 1 | 0 | 0 | 0 |
| 10 | 0 | 0 | 0 | 0 | 3 | 2 | 1 | 2 | 1 | 0 | 0 | 0 |
| 11 | 0 | 0 | 1 | 0 | 0 | 1 | 0 | 0 | 0 | 1 | 0 | 0 |
| 12 | 0 | 0 | 0 | 0 | 0 | 0 | 1 | 0 | 0 | 0 | 0 | 0 |
| 13 | 0 | 1 | 0 | 0 | 0 | 0 | 0 | 0 | 0 | 0 | 0 | 0 |
| 14 | 0 | 0 | 0 | 0 | 0 | 0 | 0 | 0 | 0 | 0 | 1 | 0 |

From the above heatmap (Table 6), the following observations can be clearly seen from the Hit movies data.

1. Class 1 has the highest tendency of Flop in the month of March and Novem-ber with 33% rate.
2. Class 2 has high tendency of Flop in the month of December with 28% rate.
3. Class 3 has high tendency of Flop in the month of February and July with 21% rate.
4. Class 4 has high tendency of Flop in the month of August with 50% rate.
5. Class 5 has a high tendency of Flop in the month of April with 31% rate.

6. Class 6 has a high tendency of Flop in the month of April with 66% rate.
7. Class 10 has high tendency of Flop in the month of February with 33% rate.
8. Class 7,8,9,11,12,13 and 14 have the least contribution in the box office Flop list.

*EDA Outcome Discussion*

From the EDA it is clearly indicated that genre has a bond with success in different months of the year. This observation provided the reasoning for the further study of Genre-Time correlation to predict movies success before its release. Figure 4 represents the hit flop rates on different months of the release.

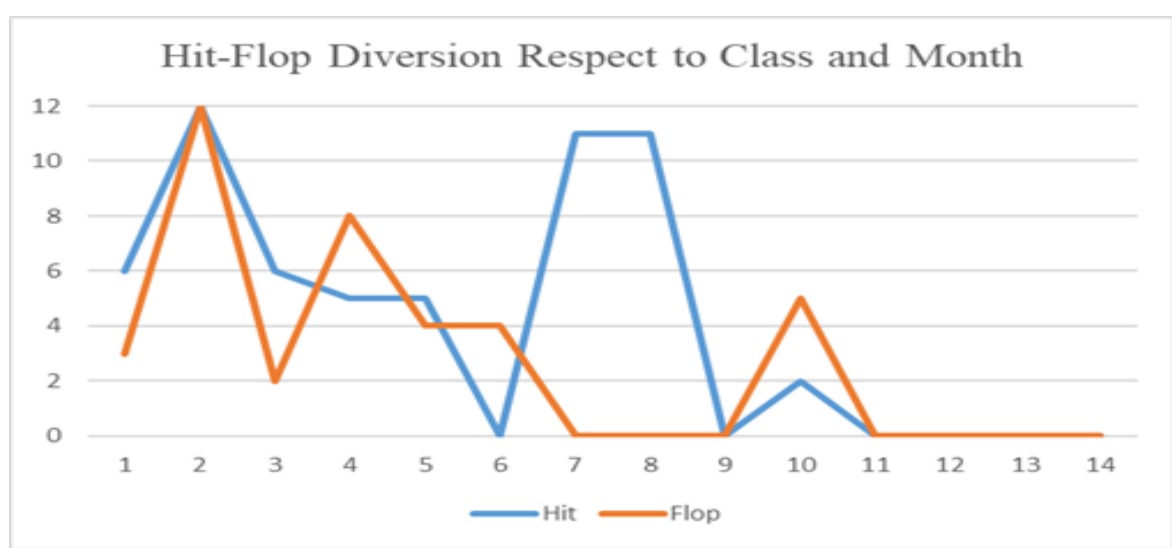

*Figure 4 Comparative representation of Hit Vs. Flop Over twelve months of the year*

*Genre-time correlation by Machine Learning Classifiers*

Following the exploration data analysis, a significant correlation was identified between movie genre and release timing. To investigate this relationship further, we developed a predictive modeling framework using supervised machine learning algorithms, aiming to assess whether a film is likely to succeed at the box office based on its genre and release month. The dataset comprises 200 top box office hits and 100 high-budget flops, compiled from IMDb, Box Office Mojo, The Numbers, and Wikipedia. Each movie instance was encoded using five primary attributes: Movie Index (Title), Genre, Release Month, Box Office Revenue, and Success Status (Hit = 1, Flop = 0). To prepare the data for modeling, categorical features such as genre were mapped to numerical class labels (e.g., Animation = 1, Superhero = 4), while release months were encoded from 1 (January) to 12 (December). Continuous features were normalized where necessary to optimize the performance of models sensitive to feature scales, such as neural networks. A total of eleven regression-based machine learning models were implemented using the WEKA platform. These included both linear and non-linear models, ranging from tree-based learners to neural networks and instance-based classifiers. Each model was evaluated using two validation strategies:

- Cross-Fold Validation with folds set to 5, 10, and 15
- Percentage Split with training-test distributions of 60–40, 70–30, and 80–20

The performance of each model was assessed using five key evaluation metrics: Correlation Coefficient (CC), Mean Absolute Error (MAE), Root Mean Squared Error (RMSE), Relative Absolute Error (RAE), and Root Relative Squared Error (RRSE). Detailed results for all eleven models are provided in Appendix A, while only the top five performers—LWT (Least Weighted Trees), Multilayer Perceptron (MLP), Random Tree, Decision Stamp, and IBk—are highlighted in the main text for clarity.

Among the evaluated models, Multilayer Perceptron and LWT consistently outperformed others, particularly under 10- and 15-fold cross-validation. MLP, a feedforward neural network trained via backpropagation, was effective in capturing non-linear dependencies between release month and genre. LWT, a tree-based algorithm, demonstrated strong performance while managing model complexity to avoid overfitting. Although WEKA does not provide explicit feature importance scores for all algorithms, internal patterns suggest that Genre and Release Month had the highest predictive weight across the top-performing models. Box Office Revenue contributed more significantly in regression contexts but had less influence on classification performance, while Success Status was used solely as the target variable.

## COMPARATIVE ANALYSIS

In the final stage of the comparisons of results, we have shortlisted the top five results from both the Cross Fold Validation approach and the Percentage Split approach. The variety of performance indicators served as the primary factor in the decision-making process, which led to the selection of these top five results (Mentioned in Table 7).

*Table 7 Cross Fold Validation Results*

Statistics of Cross Fold Validation | Visual Representation

Comparative Analysis (Cross Fold Validation - 5 Folds )

| Performance Measures | CC | MAE | RMSE | RAE | RRSE |
|---|---|---|---|---|---|
| Multilayer Perceptron | 0.9997 | 0.008 | 0.0115 | 1.5866 | 2.2881 |
| Random Forest | 0.993 | 0.0067 | 0.59 | 1.3443 | 11.7448 |
| M5Rules | 0.99 | 0.0193 | 0.0705 | 3.8479 | 14.0299 |
| LWL | 0.99 | 0.005 | 0.0707 | 0.9958 | 14.075 |
| Decision Stamp | 0.99 | 0.005 | 0.0707 | 0.9958 | 14.075 |

Comparative Analysis (Cross Fold Validation - 10 Folds)

| Performance Measures | CC | MAE | RMSE | RAE | RRSE |
|---|---|---|---|---|---|
| LWL | 1 | 0 | 0 | 0 | 0 |
| Multilayer Perceptron | 0.9997 | 0.0089 | 0.0121 | 1.7755 | 2.4075 |
| Random Forest | 0.9969 | 0.0045 | 0.397 | 0.906 | 7.904 |
| M5Rules | 0.99 | 0.0167 | 0.0696 | 3.3318 | 13.8534 |
| Random tree | 0.99 | 0.005 | 0.0707 | 0.9956 | 14.0763 |

Comparative Analysis (Cross Fold Validation - 15 Folds)

| Performance Measures | CC | MAE | RMSE | RAE | RRSE |
|---|---|---|---|---|---|
| LWL | 1 | 0 | 0 | 0 | 0 |
| Multilayer Perceptron | 0.9997 | 0.0091 | 0.012 | 1.8202 | 2.3847 |
| Random Forest | 0.9968 | 0.0044 | 0.04 | 0.8877 | 7.9821 |
| M5Rules | 0.9903 | 0.0167 | 0.0698 | 3.3377 | 13.9153 |
| Random tree | 0.99 | 0.005 | 0.707 | 0.9974 | 14.1038 |

(Cross Fold Validation - 5 Folds)

(Cross Fold Validation - 10 Folds)

(Cross Fold Validation - 15 Folds)

Table 8 has shown the comparative statistics of Percentage Split of three distributions of data simultaneously. Where LWT, Random Tree, Decision Stamp, and RepTree performed ideal on 5 and 10 folds, and LWT, Random Tree, Decision Stamp, and Ibk has shown the ideal on 15 folds, providing best results on the given dataset.

*Table 8 Percentage Split Results*

Statistics of Percentage Split | Visual Representation

| Comparative Analysis (Percentage Split 60-40) | | | | | |
|---|---|---|---|---|---|
| Performance Measures | CC | MAE | RMSE | RAE | RRSE |
| LWL | 1 | 0 | 0 | 0 | 0 |
| Random tree | 1 | 0 | 0 | 0 | 0 |
| Decision Stamp | 1 | 0 | 0 | 0 | 0 |
| Random Forest | 1 | 0.0004 | 0.0034 | 0.075 | 0.6708 |
| Rep tree | 1 | 0 | 0 | 0 | 0 |

| Comparative Analysis (Percentage Split 70-30) | | | | | |
|---|---|---|---|---|---|
| Performance Measures | CC | MAE | RMSE | RAE | RRSE |
| LWL | 1 | 0 | 0 | 0 | 0 |
| Random tree | 1 | 0 | 0 | 0 | 0 |
| Decision Stamp | 1 | 0 | 0 | 0 | 0 |
| Random Forest | 0.9999 | 0.0013 | 0.0082 | 0.2667 | 1.633 |
| Rep tree | 1 | 0 | 0 | 0 | 0 |

| Comparative Analysis (Percentage Split 80-20) | | | | | |
|---|---|---|---|---|---|
| Performance Measures | CC | MAE | RMSE | RAE | RRSE |
| LWL | 1 | 0 | 0 | 0 | 0 |
| Random tree | 1 | 0 | 0 | 0 | 0 |
| Decision Stamp | 1 | 0 | 0 | 0 | 0 |
| Ibk | 1 | 0 | 0 | 0 | 0 |
| Random Forest | 1 | 0.0005 | 0.0032 | 0.0999 | 0.632 |

## DISCUSSION

The analysis presented in this study provides compelling evidence that genre and release timing exhibit a notable correlation with a movie's box office performance. The exploratory data analysis (EDA) uncovered distinct seasonal trends in genre-based success, revealing that certain genres consistently perform better during specific months. For instance, genres like Animation and Superhero (Class 1 and Class 4) showed a high frequency of hits in June and May respectively, while genres such as Thriller and Romance (Classes 7 and 8) peaked in November. This seasonal genre preference may reflect audience behavior patterns, school holidays, or cultural events that influence movie attendance.

Conversely, the analysis of box office flops also demonstrated time-sensitive trends. For example, the Animation genre (Class 1), despite being successful in mid-year months, had a higher flop rate in March and November. Similarly, genres like Action (Class 5) and Thriller (Class 6) exhibited higher failure rates in April. These opposing trends between hits and flops across different timeframes underscore the importance of aligning genre with optimal release periods.

The EDA results provided the empirical basis for applying machine learning models to further validate and predict movie success based on genre and release timing. Using a dataset comprising the top 100 box office hits and the top 100 flops, two model validation approaches—Cross-Fold Validation and Percentage Split—were employed to evaluate predictive accuracy. The results demonstrated that models like LWT, Multilayer Perceptron, Random Tree, and Decision Stamp performed robustly, achieving strong results across five evaluation metrics (CC, MAE, RMSE, RAE, and RRSE).

Overall, the findings substantiate the hypothesis that genre and release month are significant predictors of a movie's box office outcome. Integrating temporal genre trends into predictive models not only improves forecasting accuracy but also provides actionable insights for producers, marketers, and investors. This study contributes empirical support for Genre-Time correlation and encourages its integration into strategic planning and decision-making within the movie industry.

## LIMITATION OF THE WORK

While the findings of this study highlight meaningful correlations between genre, release timing, and box office performance, several limitations must be acknowledged.

a) First, the dataset used for analysis focused solely on the top 100 box office hits and top 100 flops, which may not represent the full diversity of movie releases. This selection bias limits the generalizability of the results to the broader spectrum of films, including mid-tier or niche productions.
b) Second, the analysis does not focus on any specific regional or national film industry (e.g., Hollywood, Bollywood, or other global markets). As a result, cultural, economic, or market-specific factors that influence genre popularity and seasonal preferences are not accounted for. The findings should therefore be interpreted as general trends rather than industry-specific insights.
c) Third, the machine learning models employed in this study were based on pre-existing algorithms available in the WEKA toolkit, which, while effective, may not be optimized for the unique characteristics of movie datasets. Additionally, the limited set of input features—primarily genre and release time—may overlook other influential variables such as marketing spend, star power, or critical reviews.
d) Finally, EDA was based upon pre-cleaned data, which might have filtered out valuable insights or quirks that were found in the raw data. This may impact negatively on the precision and strength of detected patterns.

While many of these results have limitations and leave open various questions about their implications for machine learning, the study provides promising initial evidence that the relationship between genre and time in movies influences how well a movie will do, and in doing so, it paves the way for more granular, domain-driven, and rigorous research in the future.

## CONCLUSION AND FORTHCOMING GUIDELINES

The movie is a necessary factor in the material, as well as the spiritual life of all nations. As the industry evolves, there's growing interest in data-based approaches to predicting content that will do well at the box office. This research specifically investigated the relationship between movie genre and release date, which was empirically supported by EDA and positionally validated by machine learning process.

The results provide important evidence of the relationship between genres and release season, revealing that the audience appeal does depend not only on the type, but also on the time that content is distributed. Through the use of visual analytics and algorithmic modeling, this work demonstrates that genre-time correlation indeed plays a strong role in influencing the box office. The above performance metrics from both validation methods prove the effectiveness of the proposed scheme.

Although the analysis was carried out on a subset of best/worst performing movies at the box office, the results level opens new research avenues. It would be interesting in future work to use the cheaper decision model, but then to scale this up to a broader and more diverse range of data, from other regional and international industries. For instance, a cross-industry comparative study||investigating how the dynamics of genre-time dependency vary across industries—Hollywood, Bollywood, and European cinema could unveil culture or market-specific patterns in audience tastes.

Furthermore, in future studies the different distribution channels could be differentiated, particularly theatre (cinema) releases versus OTT streaming services. Since content consumption tradition is different on these mediums, the investigation results of genre-time relation on two mediums might show diversification success patterns. Furthermore, incorporating more pre-release features such as trailer ratings, marketing efforts and social media activity might help improve the prediction and make more accurate decisions for producers and distributors.

Finally, this research should help to further the increasingly accepted perspective that advanced analytics are not merely valuable, but critical to the future of media and entertainment, as well as, to finance, healthcare, retail, and other data-driven industries. By being smart with analytics, investors can ensure they make informed decisions, maximize release planning, and better guarantee commercial success in an increasingly crowded marketplace.

In conclusion, our study shows that the co-analysis of genre and release timing is predictive, and it provides a strong basis for furthering this line of investigation using larger and more complete cross-platform and cross-industry datasets.

# Appendix A

| Comparative Analysis (Cross Fold Validation) | | | | | | | | | | | | | | | |
|---|---|---|---|---|---|---|---|---|---|---|---|---|---|---|---|
| **No. of Fold** | **5** | **10** | **15** | **5** | **10** | **15** | **5** | **10** | **15** | **5** | **10** | **15** | **5** | **10** | **15** |
| **Performance Measures** | **"Correlation Coefficient" CC** | | | **"Mean Absolute Error" MAE** | | | **"Root Square Error Mean" RMSE** | | | **"Relative Absolute Error" RAE** | | | **"Root Relative square Error" RRSE** | | |
| Best Range | <1 | | | <10 | | | Close to 0 | | | <1 | | | <1 | | |
| **Process Classifiers** | **Results** | | | | | | | | | | | | | | |
| LWL | 0.99 | 1 | 1 | 0.005 | 0 | 0 | 0.0707 | 0 | 0 | 0.9958 | 0 | 0 | 14.075 | 0 | 0 |
| Multilayer Perceptron | 0.9997 | 0.9997 | 0.9997 | 0.008 | 0.0089 | 0.0091 | 0.0115 | 0.0121 | 0.012 | 1.5866 | 1.7755 | 1.8202 | 2.2881 | 2.4075 | 2.3847 |
| Random Forest | 0.993 | 0.9969 | 0.9968 | 0.0067 | 0.0045 | 0.0044 | 0.59 | 0.397 | 0.04 | 1.3443 | 0.906 | 0.8877 | 11.7448 | 7.904 | 7.9821 |
| M5Rules | 0.99 | 0.99 | 0.9903 | 0.0193 | 0.0167 | 0.0167 | 0.0705 | 0.0696 | 0.0698 | 3.8479 | 3.3318 | 3.3377 | 14.0299 | 13.8534 | 13.9153 |
| Random tree | 0.98 | 0.99 | 0.99 | 0.01 | 0.005 | 0.005 | 0.1 | 0.0707 | 0.707 | 1.9915 | 0.9956 | 0.9974 | 19.9051 | 14.0763 | 14.1038 |
| Decision Stamp | 0.99 | 0.99 | 0.99 | 0.005 | 0.005 | 0.005 | 0.0707 | 0.707 | 0.0707 | 0.9958 | 0.9956 | 0.9974 | 14.075 | 14.0763 | 14.1038 |
| Rep tree | 0.9797 | 0.99 | 0.9899 | 0.0142 | 0.006 | 0.0066 | 0.1004 | 0.0708 | 0.0709 | 2.8182 | 1.2024 | 1.3164 | 19.9887 | 14.0922 | 14.1425 |
| M5P | 0.9886 | 0.989 | 0.9891 | 0.0375 | 0.0345 | 0.0335 | 0.0764 | 0.075 | 0.0744 | 7.4718 | 6.8686 | 6.6762 | 15.2112 | 14.9238 | 14.8472 |
| Ibk | 0.98 | 0.98 | 0.98 | 0.01 | 0.01 | 0.01 | 0.1 | 0.1 | 0.1 | 1.9915 | 1.9912 | 1.9948 | 19.9051 | 19.9068 | 19.9458 |
| Gaussian Process | 0.8896 | 0.8892 | 0.8895 | 0.1855 | 0.184 | 0.1827 | 0.2299 | 0.2299 | 0.2294 | 36.9519 | 36.6385 | 36.4364 | 45.7579 | 45.7615 | 45.7593 |
| Linear Regression | 0.8896 | 0.889 | 0.8866 | 0.1753 | 0.1751 | 0.1757 | 0.2285 | 0.2291 | 0.2316 | 34.9165 | 34.8583 | 35.0497 | 45.4914 | 45.6118 | 46.199 |
| Simple Linear Regression | 0.8715 | 0.8713 | 0.8684 | 0.1861 | 0.1854 | 0.1861 | 0.2453 | 0.2455 | 0.2482 | 37.0713 | 36.9191 | 37.1274 | 48.8231 | 48.874 | 49.4986 |
| SMOreg | 0.8793 | 0.8793 | 0.8785 | 0.1508 | 0.1495 | 0.1496 | 0.2598 | 0.2623 | 0.2636 | 30.0308 | 29.7613 | 29.8436 | 51.7183 | 52.2058 | 52.5852 |

| Comparative Analysis (Percentage Split) | | | | | | | | | | | | | | | |
|---|---|---|---|---|---|---|---|---|---|---|---|---|---|---|---|
| **Train/Test Split** | **60** | **70** | **80** | **60** | **70** | **80** | **60** | **70** | **80** | **60** | **70** | **80** | **60** | **70** | **80** |
| **Performance Measures** | **"Correlation Coefficient" CC** | | | **"Mean Absolute Error" MAE** | | | **"Root Square Error Mean" RMSE** | | | **"Relative Absolute Error" RAE** | | | **"Root Relative square Error" RRSE** | | |
| Best Range | <1 | | | <10 | | | Close to 0 | | | <1 | | | <1 | | |
| **Process Classifiers** | **Results** | | | | | | | | | | | | | | |
| LWL | 1 | 1 | 1 | 0 | 0 | 0 | 0 | 0 | 0 | 0 | 0 | 0 | 0 | 0 | 0 |
| Random tree | 1 | 1 | 1 | 0 | 0 | 0 | 0 | 0 | 0 | 0 | 0 | 0 | 0 | 0 | 0 |
| Decision Stamp | 1 | 1 | 1 | 0 | 0 | 0 | 0 | 0 | 0 | 0 | 0 | 0 | 0 | 0 | 0 |
| Ibk | 0.9753 | 0.9672 | 1 | 0.0125 | 0.0167 | 0 | 0.1118 | 0.1291 | 0 | 2.5 | 3.3333 | 0 | 22.3607 | 25.8199 | 0 |
| Random Forest | 1 | 0.9999 | 1 | 0.0004 | 0.0013 | 0.0005 | 0.0034 | 0.0082 | 0.0032 | 0.075 | 0.2667 | 0.0999 | 0.6708 | 1.633 | 0.632 |
| Rep tree | 1 | 1 | 1 | 0 | 0 | 0.0114 | 0 | 0 | 0.0166 | 0 | 0 | 2.2877 | 0 | 0 | 3.3191 |
| Multilayer Perceptron | 0.3998 | 0.9998 | 0.9998 | 0.0096 | 0.0064 | 0.0133 | 0.012 | 0.0098 | 0.0169 | 1.9185 | 1.2827 | 2.6561 | 2.4036 | 1.9609 | 3.3788 |
| M5Rules | 0.9959 | 0.997 | 0.9977 | 0.0397 | 0.0364 | 0.0311 | 0.048 | 0.0439 | 0.0378 | 7.9374 | 7.2723 | 6.2126 | 9.6085 | 8.7701 | 7.5495 |
| M5P | 0.9959 | 0.997 | 0.9977 | 0.0397 | 0.0364 | 0.0311 | 0.048 | 0.0439 | 0.0378 | 7.9374 | 7.2723 | 6.2126 | 9.6085 | 8.7701 | 7.5495 |
| Gaussian Process | 0.904 | 0.9082 | 0.9056 | 0.1699 | 0.1697 | 0.1695 | 0.2148 | 0.2075 | 0.2227 | 33.9821 | 33.9456 | 33.87 | 42.9598 | 43.4939 | 44.5033 |
| Linear Regression | 0.9109 | 0.909 | 0.9078 | 0.1596 | 0.168 | 0.1599 | 0.2125 | 0.23 | 0.2288 | 31.9277 | 33.5924 | 31.6916 | 42.5007 | 46.0083 | 45.7356 |
| Simple Linear Regression | 0.8899 | 0.8836 | 0.8753 | 0.1655 | 0.1678 | 0.1656 | 0.2319 | 0.2548 | 0.2634 | 33.0916 | 33.5545 | 33.1083 | 46.3887 | 50.95 | 52.652 |
| SMOreg | 0.8991 | 0.8924 | 0.8839 | 0.1439 | 0.1565 | 0.1476 | 0.2495 | 0.2982 | 0.3152 | 28.785 | 31.3098 | 29.4943 | 49.9029 | 59.631 | 63.0046 |